\documentclass[a4paper,10pt]{article}

\usepackage[utf8]{inputenc}
\usepackage{amsmath,amsthm,amssymb,amscd,amsthm}
\usepackage{latexsym}
\usepackage{cite}
\usepackage{ifthen}
\usepackage{hyperref}
\usepackage[all,cmtip]{xy}
\usepackage{xspace}
\usepackage[labelfont={bf}]{caption}
\usepackage{color}
\usepackage{tikz}
\usepackage{tensor}
\usepackage{graphicx}
\usepackage[abs]{overpic}

\numberwithin{equation}{section}
\numberwithin{figure}{section}
\numberwithin{table}{section}

\newcommand{\sfrac}[2]{{\textstyle\frac{#1}{#2}}}
\renewcommand{\Re}{\ensuremath{\mathrm{Re}}}
\renewcommand{\Im}{\ensuremath{\mathrm{Im}}}
\newcommand{\ins}{\, \lrcorner\,}
\newcommand{\tr}{\ensuremath{\mathrm{tr}\,}}

\newcommand{\RR}{\ensuremath{\mathbb{R}}}

\newcommand{\tp}{{\!\top}}

\newcommand{\SU}{\mathrm{SU}}
\newcommand{\su}{\ensuremath{\mathfrak{su}}}
\newcommand{\SO}{\mathrm{SO}}
\newcommand{\so}{\ensuremath{\mathfrak{so}}}
\newcommand{\Sp}{\ensuremath{\mathrm{Sp}}}

\newcommand{\UU}{\ensuremath{\mathrm{U}}}

\newcommand{\g}{\ensuremath{\mathfrak{g}}\xspace}

\newcommand{\kk}{\ensuremath{\mathfrak{k}}\xspace}

\newcommand{\F}{\ensuremath{\mathcal{F}}\xspace}
\newcommand{\M}{\ensuremath{\mathcal{M}}\xspace}
\newcommand{\Mm}[1]{\ensuremath{{\mathcal{M}^{#1}}}}
\newcommand{\A}{\ensuremath{\mathcal{A}}\xspace}
\newcommand{\E}{\ensuremath{\mathcal{E}}\xspace}

\newcommand{\ii}{{\rm i\,}}
\newcommand{\dd}{{\rm d}}
\newcommand{\ee}{{\rm e}}

\newcommand{\Scs}{\ensuremath{S_\text{CS}}}
\newcommand{\Vol}{\text{Vol}}

\newcommand{\for}{\quad\text{for}\quad}
\newcommand{\und}{\qquad\text{and}\qquad}

\renewcommand{\=}{\ =\ }

\begin{document}

\begin{titlepage}
\setcounter{page}{0}
\begin{flushright}
ITP--UH--05/11\\
\end{flushright}

\vskip 2.0cm

\begin{center}

{\Large\bf Yang-Mills instantons on cones and sine-cones \\[8pt]
over nearly K\"ahler manifolds}

\vspace{12mm}

{\large 
Karl-Philip Gemmer${}^\dagger$,\ 
Olaf~Lechtenfeld${}^{\dagger\times}$,\ 
Christoph N\"olle${}^\dagger$ \ and \
Alexander~D.~Popov${}^{*}$
}
\\[8mm]
\noindent ${}^\dagger${\em
Institut f\"ur Theoretische Physik,
Leibniz Universit\"at Hannover \\
Appelstra\ss{}e 2, 30167 Hannover, Germany }
\\[8mm]
${}^\times${\em
Centre for Quantum Engineering and Space-Time Research \\
Leibniz Universit\"at Hannover \\
Welfengarten 1, 30167 Hannover, Germany }
\\[8mm]
\noindent ${}^*${\em
Bogoliubov Laboratory of Theoretical Physics, JINR\\
141980 Dubna, Moscow Region, Russia}

\vspace{12mm}

\begin{abstract}
\noindent 
We present a unified eight-dimensional approach to instanton equations on
several seven-dimensional manifolds associated to a six-dimensional 
homogeneous nearly K\"ahler manifold. The cone over the sine-cone on 
a nearly K\"ahler manifold has holonomy group Spin(7) and can be foliated 
by submanifolds with either holonomy group $G_2$, a nearly parallel 
$G_2$-structure or a cocalibrated $G_2$-structure. We show that there is 
a $G_2$-instanton on each of these seven-dimensional manifolds which gives 
rise to a Spin(7)-instanton in eight dimensions. The well-known octonionic 
instantons on $\mathbb R^7$ and $\mathbb R^8$ are contained in our 
construction as the special cases of an instanton on the cone and on 
the cone over the sine-cone, both over the six-sphere, respectively.
\end{abstract}

\end{center}
\end{titlepage}

\tableofcontents
\noindent

\section{Introduction}
 Instantons are important objects in modern field theories
 \cite{Rajaraman:1982is,Manton:2004tk}.
 Yang-Mills instantons
\cite{Belavin:1975fg}
are nonperturbative Bogomolny-Prasad-Sommerfield (BPS) configurations in four Euclidean
dimensions solving first-order anti-self-duality equations for gauge fields which imply the full Yang-Mills
equations. They play a prominent role both in mathematics
and physics, and their study has yielded many results in both areas. In this article
we discuss their higher dimensional generalization. \\

As one motivation for our study, we notice that Yang-Mills theory in more than four
dimensions naturally appears in the low-energy limit of superstring theory in the
presence of D-branes. Also, heterotic strings yield heterotic supergravity, which
contains supersymmetric Yang-Mills theory as a subsector \cite{GSW87}. Furthermore, natural BPS-type equations for gauge
fields in
dimension $d>4$, introduced in \cite{CDFN83},
also appear in heterotic superstring compactification on spacetimes $M_{10-d}\times X^d$ as
the condition of survival of at least one supersymmetry in the low-energy effective field
theory on $M_{10-d}$. These first-order instanton equations on $X^d$, which generalize the
four-dimensional ones, were considered e.g. in
 \cite{Ward84,Don85,MCS88,Tian00,DT98,Popov10},
 and some of their solutions were found in
\cite{Fairlie84,Fubini85,IvPop92,Guenaydin,Popovetal03,Haupt:2011mg,Correia09,Ivanova2009,Harland2009,Bauer2010,Harland2010}. \\

 The Yang-Mills instantons considered here can therefore be thought of as ingredients for the construction of solitons
in string theory. In heterotic string theory such solitons
were first considered in \cite{Strom90} and were interpreted as gauge 5-brane solitons, with ordinary Yang-Mills
instantons living in the four dimensions transverse to the world-volume of a flat 5-brane.
 Then, using heterotic - type I string duality, it was shown \cite{Witten95,Douglas:1995bn} 
that an instanton on $\mathbb R^4 \subset \mathbb R^{10}$ shrunk to zero size\footnote{Such singular instantons produce conical singularities in the instanton moduli
space.} corresponds to a D5-brane\footnote{More precisely, such gauge string solitons
correspond to a bound system of D1- and D5-branes \cite{Douglas:1995bn}.} in type I
string theory. \\

Not many instanton solutions are known for $d>4$. One of them is the
Spin(7)-instanton on $\mathbb R^8$ constructed by Fairlie and Nuyts and independently by
Fubini and Nicolai \cite{Fairlie84, Fubini85}. This solution was extended to a gauge
solitonic 1-brane solution of heterotic supergravity \cite{Harvey91}, the low-energy
limit of the heterotic string. A similar $G_2$-instanton \cite{IvPop92,Guenaydin} on $\mathbb R^7$ was extended to a
heterotic 2-brane soliton
independently in \cite{Ivanova:1993nu} and \cite{Guenaydin}. Due to the
connection between octonions and the groups $G_2$ and Spin(7), these gauge fields on
$\mathbb R^7$ and $\mathbb R^8$ are called octonionic instantons. In this paper we review
recent work and present some new results showing that the octonionic instantons are part
of a larger family of instantons that exist not only on Euclidean spaces but on a whole
class of conical, non-compact manifolds. This generalization of the octonionic instantons
to other spaces embeds into supergravity as well \cite{Harland11}, and it is an
interesting question whether this is true for the whole family of instantons. We leave
this to future work, however. \\

A starting point for our investigation is the fact that $\mathbb R^7$ and $\mathbb R^8$
are the metric cones over the round spheres $S^6$ and $S^7$, which carry a nearly
K\"ahler structure and a nearly parallel $G_2$-structure, respectively. Six-dimensional nearly
K\"ahler SU(3)-manifolds and seven-dimensional nearly parallel $G_2$-manifolds have weak
holonomy groups SU(3) and $G_2$, respectively. They are closely related to integrable
geometries, however, as their cones have reduced holonomy groups $G_2$ and Spin(7)
\cite{Baer93}. Furthermore, nearly K\"ahler SU(3)-structure and nearly parallel
$G_2$-structure are special among non-integrable geometries in that their natural
instanton equations imply the usual Yang-Mills equations without a torsion term
\cite{Xu,Harland11}, like it is the case for integrable $G$-structures. \\

Similar to the result that the cone has a reduced holonomy group is the observation made
in \cite{CLEYTON2002,Bilal03 } that the so-called sine-cone over a nearly K\"ahler
manifold has a nearly parallel $G_2$-structure. Like the cone, the sine-cone over a
non-spherical manifold is singular and non-complete as a Riemannian manifold. For a round
sphere, on the other hand, the cone gives Euclidean space and the sine-cone 
is again a sphere. Another important property is the fact that the cone over a sine-cone
is the same as the cylinder over a cone \cite{Boyer2007}, as illustrated for nearly
K\"ahler manifolds in Figure~\ref{diagr of geometries}. \\

Given a nearly K\"ahler manifold \M there are thus at least two different interesting
$G_2$-instanton equations in one dimension higher. First, the equation on
the cone with its integrable $G_2$-structure, and second the $G_2$-instanton equation on
the nearly parallel sine-cone. Additionally there are two
$G_2$-structures on the cylinder over the nearly K\"ahler space, but
one of those is conformally equivalent to the integrable structure on the cone, and
therefore does not give rise to a new instanton equation. \\

Of the three inequivalent $G_2$-instanton equations the two equations on a cylinder have
been studied in
a series of papers \cite{Ivanova2009,Harland2009,Harland2010,Bauer2010}, and
some explicit solutions have been found. Here we shall present a unified
approach to all these equations, by stepping up one more dimension to the
cone over the sine-cone, or equivalently the cylinder over the cone, on the
nearly K\"ahler manifold. This eight-manifold has reduced holonomy group Spin(7) and
contains all of the above-mentioned seven-manifolds as submanifolds.
Furthermore, the Spin(7)-instanton equation can be reduced to the different
$G_2$-instanton equations, and therefore all the solutions found for
particular $G_2$-structures can be considered as solutions to one single
Spin(7)-instanton equation. In
\cite{Ivanova2009,Harland2009,Harland2010,Bauer2010} a particular
SU(3)-invariant ansatz for the gauge fields has been employed, which depends
on one complex function of one complex variable when considered in eight
dimensions.\footnote{Note that more general anz\"atze were considered as
well~\cite{Harland2010,Bauer2010}.} The Spin(7)-instanton equation
turns into a first-order non-linear differential equation, with an $S_3$-symmetry that
reflects the so-called 3-symmetry
of nearly K\"ahler coset spaces \cite{Butruille2006a}. \\

Here we present the known solutions and also a new one. In addition, we show that our solutions
include the octonionic instanton on $\mathbb R^7$ \cite{IvPop92,
Guenaydin}. The gauge group for these examples can either be taken to be $G_2$, or it can be identified with the
group $G$ for a nearly K\"ahler coset space $G/H$.\footnote{There is some risk of
confusion here due to the different groups appearing. The manifolds under consideration
come equipped with a reduced weak holonomy group, which is either SU(3), $G_2$, or
Spin(7), and the six-dimensional base manifold will be chosen as a homogeneous space
$G/H$. Additionally, the gauge group of the gauge bundle plays a role.}
A generalized ansatz
with gauge group Spin(7) is discussed as well, covering the octonionic
instanton on $\mathbb R^8$~ \cite{Fairlie84, Fubini85}. \\

A review of the relevant geometric structures can be found in Section \ref{cones
section}, and the
instanton equations are presented in Section \ref{instantons section}. Section
\ref{sec_actions} contains an alternative approach to the instanton equations in terms of
a Chern-Simons action, as well as the relevant second-order equations.
Our ansatz for the gauge field is explained in
Section \ref{ansatz section}, and in \ref{subsec_solutions} we collect the known
solutions to the Spin(7)-instanton equation.
An Appendix compares the octonionic instanton on~$\RR^7$ to the one 
on the cone over~$G_2/\SU(3)$.

\section{Cones, sine-cones and cylinders over nearly K\"ahler manifolds}

\label{cones section}
We review the geometry of six-, seven- and eight-dimensional manifolds with
structure group SU(3), $G_2$ and Spin(7), respectively, discuss metric cones and
sine-cones over these manifolds and the structure they inherit from the base
manifold.\\

For any Riemannian manifold
$(\M,g)$ we define
\begin{enumerate}
 \item $C(\M)=(\RR^+ \times \M,\bar g)$ with 
       $\bar g = dr^2 + r^2 g$ as the \textit{Riemannian} or 
       \textit{metric cone} over \M,
 \item $C_s(\M)=((0,\pi) \times \M,\bar g)$ with 
       $\bar g = \dd\theta^2 +\sin^2(\theta)\,g$
       as the \textit{sine-cone} over \M and
 \item $Cyl(\M)=(\RR \times \M,\bar g)$ with 
       $\bar g = dx^2 + g$ as the \textit{cylinder} over \M.
\end{enumerate}
In the limits $\theta \rightarrow 0,\pi$ the sine-cone looks like the metric cone. \\

We can, of course, take one of these manifolds again as a base manifold for
another metric cone, sine-cone or cylinder. However, it is easy to show that for 
two of these constructions we obtain the same manifold: the cone over a sine-cone 
is the same as the cylinder over a cone, i.e.
$C(C_s(\M))=Cyl(C(\M))=(\RR\times\RR^+\times\M)$. The metric $\bar g$ of
$C(C_s(\M))$ can be rewritten in terms of coordinates  $(x,\,y)$ on $Cyl(C(\M))$ as
\begin{equation}\label{metric iterated cone}
\begin{aligned}{}
  \bar g &= \dd r^2 
             + r^2 \:(\dd\theta^2+\sin^2(\theta)\,g) \\ 
         &= \dd x^2 + \dd y^2 + y^2 \, g \:,
\end{aligned}
\end{equation}
where
 \begin{equation}\label{coordtrafo_xy+rtheta}
    (x,y) = (r\cos(\theta),
            r\sin(\theta)).
 \end{equation} 

Here we are interested in cone structures constructed over nearly K\"ahler
six-manifolds. If we normalize the nearly K\"ahler manifold such that its Einstein constant is 5, i.e. Ric $=5g$, then
 its metric cone has $G_2$-holonomy and its sine-cone admits a nearly parallel $G_2$-structure \cite{Boyer2007}. For the definition of these geometries see the following subsections.
If we did not fix the normalization of the base manifold we would have to define
the cone metric as $dr^2 + (r/r_0)^2 g$ and the sine-cone metric as $d\theta^2 +
\sin^2(\theta/\theta_0) g$ for appropriately chosen constants $r_0,\theta_0$,
depending only on the Einstein constant or scalar curvature of the base, to
obtain $G_2$-holonomy and a nearly parallel $G_2$-structure, respectively. \\

\begin{figure}[!ht] 
 \begin{align*}
  \xymatrix{
   & S^6 
     \ar[dl]_{\text{sine-cone}}
     \ar[dr]^{\text{cone}} \\
   S^7
     \ar[dr]_{\text{cone}}
   & & \RR^7 
     \ar[dl]^{\text{cylinder}} \\
   & \RR^8
  }
 \end{align*}
 \caption{Cones over $S^6$}\label{S6-cones}
\end{figure}
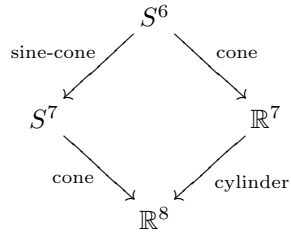

From both $G_2$-manifolds, we can construct a Spin(7)-holonomy manifold as summarized
in Figure~\ref{diagr of geometries}. The cone over a nearly parallel
$G_2$-manifold and the cylinder over a $G_2$-holonomy manifold have
Spin(7)-holonomy. For this to be true we need to normalize the nearly parallel
$G_2$-space as Ric $=6g$, which comes out right automatically if it is
constructed as a sine-cone over a nearly K\"ahler manifold normalized as above.
In general the normalization condition Ric $=(\dim \mathcal M-1)g$ on an
Einstein space $\mathcal M$ implies that its metric cone is Ricci-flat, whereas
its sine-cone is Einstein again. In the special case that we start with the nearly K\"ahler manifold $S^6$, the resulting spaces are depicted in Figure~\ref{S6-cones}.
\begin{figure}[!ht]
 \begin{align*}
  \:\\ 
  \xymatrix{
   & \text{nearly K\"ahler} 
      \ar[dl]_{\text{sine-cone }} 
      \ar[dr]^{\text{cone}}&  \\
   \text{nearly parallel }G_2 
     \ar[dr]_{\text{cone}}
   & & G_2\text{-holonomy}
     \ar[dl]^{\text{cylinder}} \\ 
   & \text{Spin(7)-holonomy} }
 \end{align*}
 \caption{Cones over nearly K\"ahler manifolds}\label{diagr of geometries}
\end{figure}
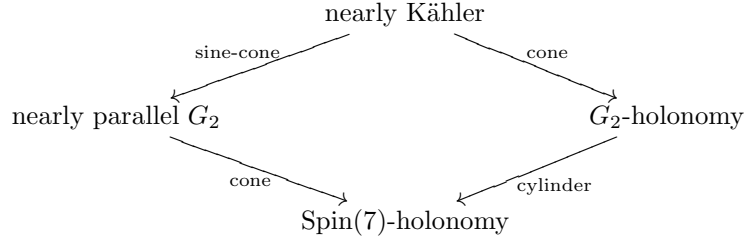

\subsection{Nearly Kähler cosets $G/H$}
\label{NK section}
Manifolds of dimension six with SU(3)-structure admit a set of canonical objects
fixed by the group SU(3), consisting of an almost complex structure $J$, a Riemannian
metric $g$, a real two-form $\omega$ and a complex three-form $\Omega$.  With
respect to $J$, the forms $\omega$ and $\Omega$ are of type (1,1) and (3,0),
respectively, and there is a compatibility condition,
$g(J\cdot,\cdot)=\omega(\cdot,\cdot)$.  With respect to the volume form $V_g$ of
$g$, $\omega$ and $\Omega$ are normalized so that
\begin{equation}
  \label{su(3)-forms normalization}
  \omega\wedge\omega\wedge\omega \= 6V_g \und
  \Omega\wedge\bar{\Omega} \= -8\ii V_g\ .
\end{equation}
A nearly K\"ahler six-manifold is an SU(3)-structure manifold such that
\begin{equation} 
\label{NK d(omega),d(Omega)}
 \dd\omega \= 3\lambda\,\Im\Omega \und 
 \dd\Omega \= 2\lambda\,\omega\wedge\omega 
\end{equation}
for some real non-zero constant $\lambda$, proportional to the square of the
scalar curvature (if $\lambda$ was zero, the manifold would be Calabi-Yau). 
Our normalization Ric $=5g$ implies that $\lambda=1$.\\

There are only four known examples of compact nearly K\"ahler six-manifolds, and
all of them are coset spaces:
\begin{equation}
\label{coset spaces}
 \begin{array}{cc}
   \SU(3)/\UU(1){\times}\UU(1)\ , & \Sp(2)/(\Sp(1){\times}\UU(1))_{\text{non-max}}\ ,\\[4pt]
   G_2/\SU(3)=S^6\ , & \SU(2)^3/\SU(2)_{\text{diag}}=S^3\times S^3\ .
 \end{array}
\end{equation}
These coset spaces $G/H$ were named 3-symmetric by Wolf and Gray, because the
subgroup $H$ is the fixed point set of an automorphism $s$ of $G$ satisfying
$s^3=\mathrm{Id}$ \cite{Wolf1967,Wolf1968,Butruille2006a}. The cosets under
consideration are all naturally-reductive, which means that there is a
decomposition $\mathfrak{g}= \mathfrak{h} \oplus \mathfrak{m}$, orthogonal with
respect to the Cartan-Killing form, where $\mathfrak g$ and $\mathfrak{h}$ are
the Lie algebras of $G$ and $H$, respectively, and $\mathfrak{m}$ is an
$\mathfrak h$-module: $[\mathfrak{h},\mathfrak{m}]\subset\mathfrak{m}$. The
tangent space of $G/H$ at a given point can be identified with $\mathfrak m$,
and in particular the almost complex structure $J$ comes from an endomorphism
$J:\mathfrak m\rightarrow \mathfrak m$. Similarly the 3-symmetry induces an
automorphism $S$ of the Lie algebra $\mathfrak{g}$ which acts trivially on
$\mathfrak h$ and is related to $J$ by
\begin{equation}\label{3 symmetry}
S|_\mathfrak{m} \= -\tfrac{1}{2} + \tfrac{\sqrt{3}}{2} J \=
\exp\left( \tfrac{2\pi}{3} J \right)\ .
\end{equation}
 The metric on $\mathfrak m$ is given by one twelfth times the Cartan-Killing form of $\mathfrak{g}$,
\begin{equation}
 g(X,Y)  \=
- \frac 1{12}\mathrm{Tr}_\mathfrak{g} ({\rm ad}(X)\circ {\rm ad}(Y))\ ,
\end{equation}
 for $X,Y\in \mathfrak m$, and due to $g$ being $H$-invariant it extends to a globally defined metric on $G/H$.
 The (1,1)-form $\omega$ is fixed by its compatibility with $g$ and
 $J$, and $\Omega$ is the unique suitably normalized $G$-invariant (3,0)-form.\\

In calculations, it is useful to choose a basis $\{I_A\}$ for the Lie algebra $\mathfrak{g}$.  We do so in such a way
that $I_a$ for $a=1,\dots,6$ form a basis for $\mathfrak{m}$ and $I_i$ for $i=7,\dots,\mathrm{dim}(G)$ yield a basis for
$\mathfrak{h}$. Furthermore, we impose a suitable normalization for the
structure constants $f_{AB}^C$:
\begin{equation}
[I_A,I_B]\=f_{AB}^CI_C \qquad\textrm{with}\qquad 
f_{AC}^Df_{DB}^C\=12\delta_{AB}\ .
\end{equation}
 Then $f_{ABC}:=f_{AB}^D\delta_{DC}$ is totally antisymmetric.  
The reductive property of the coset means that the structure constants $f_{aij}$ vanish. Then the 3-symmetry implies useful identities involving $\omega$: notably, the tensor
\begin{equation}
\label{f identity 1}
\tilde{f}_{abc}\ :=\ f_{abd}\omega_{dc}
\end{equation}
is totally antisymmetric; furthermore,
\begin{equation}
\label{f identity 2}
\omega_{ac}f_{cbi} \= \omega_{bc}f_{cai}\und 
\omega_{ab} f_{abi} \= 0\:,
\end{equation} 
 which tell us that the endomorphisms $I_a \mapsto f^b_{ia} I_b $ of $\mathfrak m$ are contained in the $\mathfrak{su}(3)$-subalgebra of $\mathfrak{so}(\mathfrak m) \cong \mathfrak{so}(6)$ defined by the almost complex structure. \\

The metric and almost complex structure on $\mathfrak{m}$ lift to a
$G$-invariant metric and almost complex structure on $G/H$.  Local expressions
for these can be obtained by introducing an orthonormal frame as follows.  The
basis elements $I_A$ of the Lie algebra $\mathfrak{g}$ can be represented by
left-invariant vector fields $\hat E_A$ on the Lie group $G$, and the dual basis
$\hat{e}^A$ is a set of left-invariant one-forms.  The space $G/H$ consists of
left cosets $gH$, and the natural projection $g\mapsto gH$ is denoted by
$\pi:G\rightarrow G/H$.  Over a contractible open subset $U$ of $G/H$, one can
choose a map $L:U\rightarrow G$ such that $\pi\circ L$ is the identity (in other
words, $L$ is a local section of the principal bundle $G\rightarrow G/H$).  The
pull-backs of $\hat{e}^A$ under $L$ are denoted by $e^A$.  In particular, $e^a$ form
an orthonormal frame for $T^* (G/H)$ over $U$ (where again $a=1,\dots 6$), and
we can write $e^i=e^i_ae^a$ with real functions $e^i_a$.  The dual frame for
$T(G/H)$ will be denoted by $E_a$.  The forms $e^A$ obey the Maurer-Cartan equations,
\begin{equation}
\begin{aligned}
\label{MC}
\dd e^a &\=\ -f_{ib}^a\;e^i\wedge e^b\ -\ \tfrac12 f_{bc}^a\;e^b\wedge e^c\ , \\
\dd e^i &\=\!-\tfrac12 f_{bc}^i\,e^b\wedge e^c\ -\ \tfrac12 f_{jk}^i\,e^j\wedge e^k\ .
\end{aligned}
\end{equation}
Since all the connections we will consider are invariant under some action of $G$,
it suffices to do calculations just over the subset $U$. Local expressions for the $G$-invariant metric, almost complex
structure, and
nearly K\"ahler form on $G/H$ are then
\begin{eqnarray}
 g \=  \delta_{ab}e^a e^b\ , \qquad
 J \= {J^a}_b E_a e^b  \und
 \omega \= \tfrac{1}{2}\omega_{ab} e^a\wedge e^b\:,
\end{eqnarray}
 where in fact $\delta_{ac} J^c_b = \omega_{ab}$.
 One can also obtain a local expression for the (3,0)-form $\Omega$.  From
(\ref{MC}) it follows that
\begin{equation}
\dd\omega \= -\tfrac{1}{2} \tilde{f}_{abc}\, e^a\wedge e^b\wedge e^c
\und
\ast \dd\omega \= \tfrac{1}{2} f_{abc}\, e^a\wedge e^b\wedge e^c\ .
\end{equation}
As we have $\dd\omega= 3\Im\Omega$ it must be that
\begin{equation}
\label{Omega}
\mathrm{Im}\,\Omega \=
-\tfrac{1}{6}\tilde{f}_{abc}\,e^a\wedge e^b\wedge e^c
\und
\mathrm{Re}\,\Omega \=
-\tfrac{1}{6}f_{abc}\, e^a\wedge e^b\wedge e^c\ .
\end{equation}

\subsection{$G_2$-structures}
Consider a seven-dimensional manifold with the structure group of its tangent bundle
contained in $G_2$. The $G_2$-invariant objects we have at our disposal are the
Riemannian metric and a real three-form $\Psi$.  Locally, it is always possible to choose an
orthonormal coframe $\{e^1,\ldots,e^7\}$ such that $\Psi$ can be expressed as
\begin{align}
 \Psi = \tfrac{1}{3!}\,f^{\mathbb O}_{abc}\,e^a\wedge e^b\wedge e^c
 \qquad\text{with}\quad a,b,c=1,\ldots,7 \:,
\end{align}
with $f^{\mathbb O}_{abc}$ being the octonionic structure constants.  If $\Psi$
is closed and coclosed, the manifold has holonomy contained in $G_2$.  
\paragraph{Cones.} For nearly parallel $G_2$-manifolds, by definition $\Psi$ satisfies
\begin{align}
 \dd\Psi = \gamma *\!\Psi \qquad (\text{implying } \dd *\!\Psi=0)
\end{align}
for some constant $\gamma \in \RR$. The normalization Ric $=6g$ implies
$\gamma=\pm 4$. The $G_2$-manifolds that are of interest for us are cones and
sine-cones $(\Mm7,\bar g)$
on a nearly K\"ahler manifold $(\Mm6,g)$.  First, we consider the sine-cone: We
can use the two- and three-forms $\omega$ and $\Omega$ on the base manifold to
define a three-form
\begin{align}
\label{Psi sine-cone}
 \Psi^{sc} &= \sin^2(\theta)\,\omega\wedge \dd \theta
          + \sin^3(\theta)\,
                \Im(\ee^{\ii \theta} \Omega ) 
\end{align}
on the sine-cone over \Mm6.
It satisfies $\dd \Psi = 4 *\!\Psi$ and induces 
a nearly parallel $G_2$-structure on the sine-cone.
On the metric cone with a radial variable $y$ a three-form can be defined as
\begin{align}
\label{Psi cone}
 \Psi^c &= y^2\,\omega\wedge \dd y + y^3\, \Im\Omega.
\end{align}
It is closed and coclosed, reflecting the fact that the cone
on a nearly K\"ahler manifold has $G_2$-holonomy.  As mentioned in Section
\ref{cones section} the metric cone can be regarded as the limit of the
sine-cone for $\theta \rightarrow 0,\pi$.  Equivalently, the three-form on the
cone \eqref{Psi cone} can be obtained by considering the corresponding form
\eqref{Psi sine-cone} on the sine-cone in this limit.

\paragraph{Cylinders.}
There are two interesting $G_2$-structures on the cylinder $\mathbb R\times \M$, with its metric $\bar g=d\tau ^2 + g$. They are
 \begin{equation}
\begin{aligned}{}
   \Psi^1 &= \omega \wedge d\tau +  \mathrm{Im}\, \Omega\:,\\
   \Psi^2 &= \omega \wedge d\tau -  \mathrm{Re}\, \Omega\:.
  \end{aligned}  
\end{equation} 
 Note that the cylinder metric is conformally equivalent to the cone metric, under the substitution $y=e^\tau $. Under this conformal equivalence the 3-form $\Psi^1$ gets identified with the 3-form \eqref{Psi cone}, defining the parallel $G_2$-structure on the cone. Therefore we call the cylinder equipped with $\Psi^1$ \emph{conformally parallel}. The 3-form $\Psi^ 2$ on the other hand is coclosed but neither closed nor nearly parallel, and such general $G_2$-structures are called \emph{cocalibrated}. Manifolds with cocalibrated $G_2$-structures admit a 
 compatible connection with totally skew-symmetric torsion \cite{Friedrich02}, which makes them promising candidates for supergravity backgrounds. 
 For $\theta$ close to $\pi/2$ the sine-cone looks like a cocalibrated cylinder.

\subsection{Spin(7)-holonomy from $G_2$-structure manifolds} 
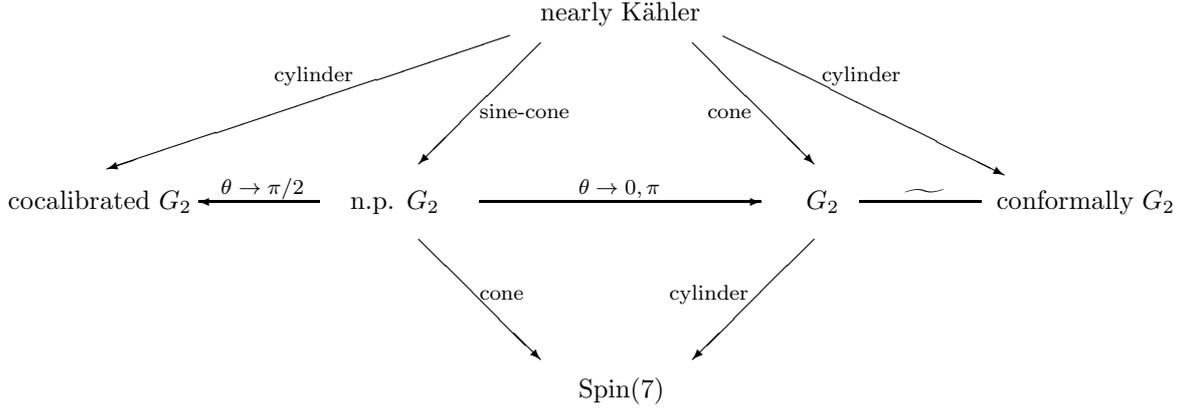
\begin{figure}\quad
\setlength{\unitlength}{1mm}
\begin{picture}(160,50)
\put(70,50){nearly K\"ahler}
\put(70,47){\vector(-1,-1){16}}\put(62,37){\footnotesize  sine-cone}
\put(45,25){n.p. $G_2$}
\put(90,47){\vector(1,-1){16}}\put(92,37){\footnotesize  cone}
\put(105,25){$G_2$}
\put(62,26){\vector(1,0){37}}
\put(75,27){\footnotesize $\theta \rightarrow 0,\pi$}
\put(54,21){\vector(1,-1){16}}\put(62,13){\footnotesize cone}
\put(106,21){\vector(-1,-1){16}}\put(87,13){\footnotesize cylinder}
\put(75,0){Spin(7)}
\put(28,27){\footnotesize $\theta \rightarrow \pi/2$}
\put(41,26){\vector(-1,0){16}}
\put(0,25){cocalibrated $G_2$}
\put(112,26){\line(1,0){16}}\put(117,25){$\widetilde{\qquad}$}
\put(130,25){conformally $G_2$}
\put(66,48){\vector(-3,-1){53}}\put(35,42){\footnotesize cylinder}
\put(94,48){\vector(2,-1){37}}\put(107,42){\footnotesize cylinder}
\end{picture}
\caption{Metric cones, sine-cones and cylinders over nearly K\"ahler manifolds. The vertical
position of a manifold gives its dimension, from six at the top to eight at the
bottom. Here $G_2$ and Spin(7) denote holonomy groups of manifolds, and n.p.
$G_2$ is a nearly parallel $G_2$-manifold. The cylinder over a nearly K\"ahler
manifold is conformally equivalent to the cone, and therefore carries a
conformally parallel $G_2$-structure. There is also a cocalibrated $G_2$-structure on the cylinder, however, which is why it occurs
twice. All spaces in this diagram are submanifolds of the Spin(7)-manifold
$\Mm8$, and the Spin(7)-structure on \Mm8 induces the cocalibrated $G_2$-structure on the cylinder, rather than the conformally parallel one.}\label{diagram2}
\end{figure}

Consider now an eight-dimensional manifold \Mm8. It is called a
Spin(7)-manifold if it comes equipped with a Riemannian metric $g$ and a
closed self-dual four-form $\Sigma$. As illustrated in Figure~\ref{diagr of
geometries} there are two equivalent possibilities to obtain a Spin(7)-holonomy
manifold $(\Mm8,\tilde g)$ by a cone construction on a nearly K\"ahler manifold
\Mm6: the metric cone over the nearly parallel $G_2$-manifold $C_s(\Mm6)$ or the
cylinder over the $G_2$-holonomy manifold $C(\Mm6)$. They give rise to two
convenient sets of coordinates, $(x,y) \in \mathbb R\times \mathbb R^+$ and $(r,\theta)\in \mathbb R^+\times (0,\pi)$, related as in \eqref{coordtrafo_xy+rtheta}. In both cases, we can use the $G_2$-invariant three-form $\Psi$ on the base $(\Mm7,\bar g)$ to define a four-form
 \begin{equation}
\begin{aligned}{}
    \Sigma &= \dd x \wedge \Psi^c \: + *_7 \Psi^c \\
 & = r^3 \dd r \wedge \Psi^{sc} \:+ r^4 *_7\!\Psi^{sc}\:,
\end{aligned}
 \end{equation} 
  where the metric on \Mm8 is 
 \begin{equation}
   \tilde g = \dd x^2 + \bar g^{c}= \dd r^2 + r^2 \bar g^{sc}\:,
 \end{equation} 
with $\bar g^c$ and $\bar g^{sc}$ the cone and sine-cone metric, respectively. 
A good way to see that the two four-forms coincide is by noting that they can be written as
\begin{align}
 \Sigma = \tfrac{1}{2}\tilde\omega\wedge\tilde\omega - \Re\,\tilde\Omega \:,
\end{align}
where
\begin{subequations}
 \label{su4-structure}
 \begin{alignat}{2}
  \tilde\omega &= y^2\, \omega + \dd x \wedge \dd y \qquad \und & 
  \tilde\Omega &= y^3 \,\Omega\wedge(\dd y - \ii\dd x)
 \intertext{in terms of the canonical coordinates on $Cyl(C(\Mm6))$, or
              in coordinates of $C(C_s(\Mm6))$:}
 \tilde\omega &= r^2\,\sin^2(\theta)\,\omega 
                  + r\,\dd r\wedge\dd \theta \und
 & \tilde\Omega &= -\ii r^3\,\sin^3(\theta)\,
                    \Omega\wedge\dd\big(\ee^{\ii \theta }r\big) \:.
\end{alignat}
\end{subequations}
 Although we have not found a simple way to construct the 8-dimensional manifold $Cyl(C(\Mm6))$ from the cylinder $Cyl(\Mm6)$, the latter one is contained in it as the submanifold $y= y_0 = $ const. The relation between the different spaces is illustrated in Figure~\ref{diagram2}.

\section{Instantons and submanifolds}\label{instantons section}
 We consider the instanton equation on the geometries discussed in the previous section. Let $\E$ be a principal $G$-bundle over a $n$-manifold \M and
\A a connection on \E, with curvature 2-form $\F:= \dd\A+\A\wedge\A$.
The instanton equation can be written as 
\begin{align}
\label{H-anti-self-dualtity}
 *\F = - \Xi \wedge \F
\end{align}
for a $(n{-}4)$-form $\Xi$ on $\M$. More precisely,
\begin{align}
\label{instanton equations - Xi}
 \Xi = \Bigg\{\begin{array}{cc}
       \omega &\for n=6 \:, \\
       \Psi   &\for n=7 \:, \\
       \Sigma &\for n=8 \:,
     \end{array}
\end{align}
where $\Psi$ and $\Sigma$ are the forms defining the $G_2$- and Spin(7)-structure
on a seven- or eight-manifold and $\omega$ is the
nearly K\"ahler (1,1)-form in six dimensions. By differentiating \eqref{H-anti-self-dualtity} it follows that instanton
solutions satisfy a Yang-Mills equation with torsion
\begin{equation}
 \label{YM with torsion}
 D_\A *\!\F + *{\cal H} \wedge \F  =   0 \:,
\end{equation} 
where the torsion three-form $\mathcal H$ is defined via $\Xi$ as
\begin{equation}
 *{\cal H }:= \dd \Xi 
\end{equation}
and
\begin{equation}
D_\A *\!\F:= \dd *\!\F + \A\wedge *\F +(-1)^{n-1} *\!\F \wedge \A
\end{equation}
in $n$ dimensions. On the manifolds of interest to us, however, the instanton
equation also implies
the Yang-Mills equation without torsion. For manifolds with $G_2$- or
Spin(7)-holonomy this is obvious since we have $\dd \Xi=0$ in these cases. 
Furthermore, it can be shown that the second term in \eqref{YM with torsion}
vanishes if \F is an instanton on a nearly K\"ahler or nearly parallel
$G_2$-manifold. In fact, the SU(3)-instanton equation on a nearly K\"ahler manifold can also be written as \cite{Xu}
 \begin{equation}
   \F\wedge \Omega =0
 \end{equation} 
 and is equivalent to the Hermitian-Yang-Mills equation $\F^{(2,0)} =  \F\lrcorner\, \omega =0$. The instanton equation on a $G_2$-manifold on the other hand has the alternative formulation
 \begin{equation}
    \F\lrcorner \, \Psi =0\:,
 \end{equation} 
 so that in both cases instantons satisfy $\F \wedge \dd\Xi =0$.\\

The instanton equations \eqref{H-anti-self-dualtity} have a natural
interpretation in terms of the Lie algebra \kk of the structure group of the
tangent bundle $T\M$.  The space of two-forms on \M at a given point is isomorphic to the Lie
algebra \so($n$) for $\dim \M=n$.  Thus, if \g denotes the Lie algebra of the
gauge group $G$, \F assumes values in
$\Lambda^2T^*\M\otimes\g\cong\so(n)\otimes\g$.  From this point of view, the
instanton equations \eqref{H-anti-self-dualtity} are equivalent to the condition
that \F assumes values in $\kk\otimes\g\subset\so(n)\otimes\g$ with
$\kk=\su(3)$, $\g_2$ or $\mathfrak{spin}(7)$ in dimension six, seven and eight, 
respectively.\\

In the following we will consider the Spin(7)-instanton equation on 
$\Mm8=Cyl(C(\Mm6))=C(C_s(\Mm6))$ with \Mm6 being a nearly K\"ahler coset space. Of particular interest is the 
reduction of this instanton equation to the submanifolds $C(\Mm6),\ C_s(\Mm6)$ and $Cyl(\Mm6)$
as well as and its relation to the $G_2$-instanton equations on these submanifolds. 
  Finally, we consider
the reduction of these equations to \Mm6 and compare it to the
SU(3)-instanton equation.\\

Let \A be a connection on \Mm8 with curvature \F satisfying the
Spin(7)-instanton equation and select an oriented submanifold \Mm7 in \Mm8. 
Later we will specify \Mm7 to be $C(\Mm6)$, $C_s(\Mm6)$ or $Cyl(\Mm6)$.  We denote
the one-form dual to the unit normal vector of \Mm7 by $\nu_8$.  A
generic $p$-form $\alpha$ on \Mm8 can be decomposed as
\begin{align}
 \alpha = \beta_1 + \nu_8\wedge\beta_2
\end{align}
with $\beta_1\in\Gamma(\Lambda^pT\Mm7)$ and
$\beta_2\in\Gamma(\Lambda^{p-1}T\Mm7)$.  Applying this to the Spin(7)-instanton
equation 
\begin{equation}
 *_8 \F=-\Sigma\wedge\F
\end{equation} 
 restricted to \Mm7, it yields
\begin{subequations}
\begin{align}
\label{G2-inst eq from Spin7}
 *_7 \F &= -\Psi\wedge\F + (*_7\Psi) \wedge (\nu_8 \ins \F) \:,\\[2pt]
\label{G2-inst eq form Spin7 v2}
 \nu_8\ins\F &= \F\ins\Psi \:.
\end{align}
\end{subequations}
 In fact these two equations are equivalent, which can be proven by decomposing
 $\F$ according to the splitting $\mathfrak{so}(7)= \mathfrak g_2
 \oplus \mathfrak m'$ into its $\mathfrak g_2$ and $\mathfrak m'$-components, and
 using the fact that the $\mathfrak g_2$-instanton operator $\F\mapsto *(\Psi
 \wedge \F)$ has eigenvalues $-1$ on $\mathfrak g_2$ and $2$ on $\mathfrak m'$.
 Additionally one needs the property that $\F\in \mathfrak g_2$ is equivalent to
 $ \F\lrcorner \Psi =0$. Like the $G_2$-instanton equation, 
 \eqref{G2-inst eq form Spin7 v2} does not restrict  the $\mathfrak g_2$-part of
 \F, but it allows for a non-vanishing $\mathfrak m'$-component as well. In
 particular, for $\nu_8\ins\F=0$ the Spin(7)-instanton equation on \Mm8 becomes
 equivalent to the $G_2$-instanton equation on the submanifold \Mm7.  \\

Furthermore, we can reduce \eqref{G2-inst eq from Spin7} back to
\Mm6.  We can do this equivalently from the cone $C(\Mm6)$ or the sine-cone
$C_s(\Mm6)$.  Analogously to the last paragraph, we denote the one-form dual to
the unit normal vector to \Mm6 by $\nu_7$ and split \eqref{G2-inst eq from
Spin7} into three equations:
\begin{subequations}
\begin{align}
\label{SU(3)-inst eq from Spin7}
 *_6 \F &= -\omega\wedge\F 
           -\sfrac{1}{2} \omega\wedge\omega\,(\nu_7\ins\nu_8\ins\F) 
           -\Im\left( \ee^{-\ii\sigma} 
               \Omega\wedge(\nu_7\ins\F + \ii \nu_8\ins\F) \right) \:,\\[2pt]
\label{SU(3)-inst eq - 2,0 part}
  \F\ins\Omega
    &= -\ii \ee^{-\ii\sigma}(\mathrm{Id}-\ii J)(\nu_7\ins\F +\ii\nu_8\ins\F) \:, \\[3pt]
\label{SU(3)-inst eq - omega part}
   3\F\ins \omega &= \nu_7\ins \nu_8\ins F\: ,
\end{align}
\end{subequations}
 where $\sigma$ is defined as follows. If we follow the path
\begin{align*}
 Cyl(C(\Mm6)) \rightarrow C(\Mm6) \rightarrow \Mm6 
 \qquad\text{we set}\quad\sigma=0 \:,
\intertext{and if we reduce to \Mm6 via}
 C(C_s(\Mm6))\rightarrow C_s(\Mm6)\rightarrow\Mm6
 \qquad\text{we set}\quad\sigma=\theta \:,
\end{align*} 
with $\theta$ being the extra coordinate on the sine-cone. However, \eqref{SU(3)-inst eq from Spin7} already implies
\eqref{SU(3)-inst eq - 2,0 part} and \eqref{SU(3)-inst eq - omega part}. 
Of course, both
equations could also be derived from \eqref{G2-inst eq form Spin7 v2}. We may decompose $\F$ as
\begin{align}
 \F = \F^{2,0} + \F^{0,2} + \mathring\F^{1,1} + \F^\omega\,\omega \:,
\end{align}
where $\F^{2,0}$ and $\F^{0,2}=\overline{\F^{2,0}}$ are (2,0)- and (0,2)-forms
with respect to the almost complex structure $J$ while $\mathring\F^{1,1}$ is a (1,1)-form with zero
$\omega$-trace.  Equivalently, we can decompose \eqref{SU(3)-inst eq from Spin7}
into two equations determining the (2,0)$\oplus(0,2)$- and $\omega$-part of
\F. The (1,1)-part orthogonal to $\omega$ is unrestricted, as in the usual SU(3)-instanton equation.
The (2,0)$\oplus(0,2)$ component of \F is determined by 
\eqref{SU(3)-inst eq - 2,0 part}, and \eqref{SU(3)-inst eq - omega part}
governs the $\omega$-part.

\section{Yang-Mills actions and Chern-Simons flows}
\label{sec_actions}
In this section we discuss an action principle that leads to the torsionful Yang-Mills equation \eqref{YM with torsion}. The action consists of the ordinary Yang-Mills action plus an additional Chern-Simons term. Moreover, we clarify the relation between the pure Chern-Simons-type action on nearly K\"ahler or nearly parallel $G_2$-manifolds and the respective instanton equations.\\

Consider the Yang-Mills action with torsion on a manifold \M,
\begin{alignat}{2}
\label{YM-action w torsion}
 S &= S_\text{YM} + S_\text{CS} 
\intertext{with}
  S_\text{YM} &= \tfrac12 \int_\M \tr(\F\wedge *\F) 
  \qquad\textrm{and}\qquad
& S_\text{CS} &= \tfrac12 \int_\M \tr(\F\wedge\F\wedge\Xi)
\end{alignat}
where $\Xi$ is defined as in \eqref{instanton equations - Xi}. 
The variation of $S$ yields the Yang-Mills equation with torsion
\begin{align}
\label{YM eq with torsion - after action}
  D_\A *\!\F + \dd\Xi \wedge \F  = 0  \:,
\end{align}
which we also obtained in the previous section by differentiating the instanton equation \eqref{H-anti-self-dualtity}.\\

On manifolds with $G_2$- and Spin(7)-holonomy the second term vanishes since $\Xi$ is closed in these cases. As discussed in the preceding section, for SU(3)-instantons on nearly K\"ahler manifolds and $G_2$-instantons on nearly parallel $G_2$-manifolds also both terms in \eqref{YM eq with torsion - after action} vanish separately.  Thus, it is also enlightening to study the Chern-Simons-type action $\Scs$ by itself, and we will consider the equation of motion obtained from $\Scs$ on a six-dimensional nearly K\"ahler manifold \Mm6 and on the sine-cone $\Mm7=C(\Mm6)$ over a nearly K\"ahler manifold.\\

\paragraph{Chern-Simons flow on nearly Kähler manifolds.}
On a nearly K\"ahler manifold \Mm6 we have $\Xi=\omega$, and the equation of motion resulting from $S_\text{CS}$ is
\begin{align}
 \dd\omega\wedge\F=0 \:,
\end{align}
 which is exactly the SU(3)-instanton equation.
Furthermore, we can consider the gradient flow equation for the Chern-Simons-type action on $\Mm6$,
\begin{align}
\label{nk flow on cyl}
 \frac{\dd \A}{\dd \tau} &= *(\F\wedge\dd\omega) 
\end{align}
where $\tau$ is a real parameter.  In \cite{Harland2009} it has been shown that \eqref{nk flow on cyl} is equivalent to the $G_2$-instanton equation on the cylinder with the conformally parallel $G_2$-structure. Due to the conformal invariance of the instanton equation this is also equivalent to the $G_2$-instanton equation on the cone $C(\M^6)$. Moreover, it has been shown, that the Hamiltonian flow equation
\begin{align}
 J\,\frac{\dd \A}{\dd \tau} &= *(\F\wedge\dd\omega)
\end{align}
is equivalent to the $G_2$-instanton equation on the cocalibrated cylinder.\\

\paragraph{Chern-Simons flow on nearly parallel $G_2$-manifolds.}
On the sine-cone over \Mm6 ($\Xi=\Psi$) we can rewrite the equation of motion for $S_\text{CS}$ as
\begin{align}
  0=\dd\Psi\wedge\F =4 *\!\Psi\wedge\F \:,
\end{align}
i.e. the $G_2$-instanton equation. Furthermore, the gradient flow equation for the Chern-Simons-type action $S_\text{CS}$ on $C_s(\Mm6)$ is
\begin{align}
\label{Flow on cyl}
 \frac{\dd \A}{\dd \tau} &= *(\F\wedge\dd\Psi) 
                       = 4\,\F\ins\Psi 
\end{align}
 for a real parameter $\tau$.
Equation \eqref{Flow on cyl} is equivalent to the Spin(7)-instanton equation \eqref{G2-inst eq form Spin7 v2} on $Cyl(\Mm7)$ and, due to the conformal invariance of the instanton equation, also to the Spin(7)-instanton equation on the cone $C(\M^7)$.

\section{Explicit ansätze for the connection}
\label{ansatz section}

 A number of instanton solutions on nearly K\"ahler, $G_2$- and Spin(7)-structure
manifolds \cite{Harland2009,Harland2010,Ivanova2009,Bauer2010} are already known
in the literature. Here we give a brief review of some of these solutions on
cones, sine-cones and cylinders. Our starting point is the so-called canonical
connection, which exists on every nearly K\"ahler manifold and can be written as
$\hat \A = e^iI_i$ on coset spaces, using the notation of Section \ref{NK section}.
It has holonomy SU(3) and satisfies the SU(3)-instanton equation.  Instantons
with gauge groups $G$ or $G_2$ will be found by adding further terms to $\hat \A$. 
We will also sketch a similar construction for gauge group Spin(7), starting from the canonical $G_2$-connection on the sine-cone.
 
\subsection{Reduced instanton equations for gauge group $G$}
\label{sec gauge group G}
A generic $G$-invariant connection on $\Mm6=G/H$ with gauge group equal to $G$ can be written as
\begin{align}
 \A = e^i I_i +e^a  \Phi_{ab} I_b 
\end{align}
where $ \Phi$ must be $H$-invariant. We choose
\begin{align}
\label{ansatz1 for A}
 \Phi_{ab} = \phi_1 \, \delta_{ab} + \phi_2 \, \omega_{ab} 
\end{align}
with $\phi_{1,2} \in \RR$. This is the most general possibility for $G_2/$SU(3); for other coset spaces further
$G$-invariant gauge fields exist \cite{Harland2009, Bauer2010}. The curvature of this connection is given by
\begin{align}
\label{F NK}
\F &= \tfrac{1}{2} \left[ f_{aci} (\Phi^\tp\Phi-\mathrm{Id})_{cb}\,I_i
       + f_{abc} ( -\Phi+(\Phi^\tp)^2)_{cd}\,I_d \right] e^a\wedge e^b\:.
\end{align}
For this ansatz, the SU(3)-instanton equation is equivalent to
\begin{align}
\label{su3 instanton eq for phi}
 -\phi+\overline\phi^2 = 0 \qquad\text{where}\qquad 
  \phi := \phi_1 + \ii\phi_2\:.
\end{align}
This equation has four solutions: 
$\phi=0$, $\phi=1$ and $\phi=\exp\left(\pm\ii\frac{2\pi}{3}\right)$. Except for $\phi=0$ the curvature vanishes at these points.
If we consider $\phi$ as a function of the additional coordinate $r,\tau $ or $\theta$ on cone, cylinder or sine-cone, respectively, then the curvature \eqref{F NK} acquires additional contributions of the form 
\begin{equation}
 \partial_\tau \phi_1\, d\tau \wedge e^a I_a + \partial_\tau \phi_2\, d\tau \wedge e^a \omega_{ab} I_b\:,
\end{equation} 
 and the $G_2$-instanton equation yields a differential equation for $\phi$:
\begin{subequations}
\label{real de for phi}
\begin{align}
 \label{G2-inst eq for phi on cyl1}
  \frac 12 \frac{\dd\phi}{\dd \tau} &= -\phi + \overline\phi^2  
 \qquad\text{{(conformally parallel cylinder)}},\\
  \label{G2-inst eq for phi on cyl2}
  \frac {\ii} 2 \frac{\dd\phi}{\dd \tau}& = -\phi + \overline\phi^2  
 \qquad\text{{(cocalibrated cylinder)}} ,\\
\label{G2-inst eq for phi on cone}
  \frac y2\,\frac{\dd\phi}{\dd y} &= -\phi + \overline\phi^2 
   \qquad\text{{(metric cone)}}\, , \\
\label{G2-inst eq for phi on sine-cone}
 \frac 12 \sin(\theta)\,\ee^{-\ii\theta}\,\frac{\dd\phi}{\dd
 \theta} &= -\phi + \overline\phi^2  
      \qquad\text{{(sine-cone)}} .
\end{align}
\end{subequations}
 As explained before, the cone equation \eqref{G2-inst eq for phi on cone} is equivalent to the standard instanton
 equation \eqref{G2-inst eq for phi on cyl1} on the cylinder, via the substitution $y=\ee^\tau$. It can be considered as a gradient flow equation, whereas the cylinder equation \eqref{G2-inst eq for phi on cyl2} admits an interpretation as a Hamiltonian flow equation
\cite{Harland2009}. 
The equations \eqref{real de for phi} appear more naturally from an eight-dimensional point of view, if we consider
$\Mm8 = Cyl(C(\Mm6))=C(C_s(\Mm6))$ with the metric 
\begin{equation}
 g^{(8)}=\dd x^2+\dd
y^2+y^2\,g^{(6)}\:.
\end{equation} 
Choose \A as above with $\phi$ being a complex function of $x$ and $y$.
 By defining a complex coordinate $z=y-\ii x$ on
$\RR^ +\times\RR$, the Spin(7)-instanton equation for this ansatz can be written
as
\begin{align}
\label{cplx de for phi}
  \,\text{Re} (z) \frac{\dd\phi}{\dd z} = -\phi+\overline\phi^2\:.
\end{align}
This equation reduces to the differential equation for the cone, the sine-cone or the cocalibrated cylinder if we restrict it to a particular path in the complex half-plane spanned by $z$. 
One chooses
\begin{itemize}
  \item $z= y+ \ii x_0$ for some constant $x_0\in\RR$ to obtain the metric cone and \eqref{G2-inst eq for phi on cone},
  \item $z=-\ii r_0 \ee^{\ii \theta}$ for some constant $r_0>0$ to obtain the sine-cone and \eqref{G2-inst eq
for phi on sine-cone},
  \item $z=y_0  - \ii \tau $ for some constant $y_0\in\RR^+$ to obtain the cocalibrated cylinder with metric $\bar g = d\tau^2 + y_0^2 g^{(6)}.$ The instanton equation reads 
 \begin{equation}\label{coccyl:insteq2}
   \frac {\ii y_0} 2 \frac{\dd\phi}{\dd \tau} = -\phi + \overline\phi^2 , 
 \end{equation} 
 which is a slight generalization of \eqref{G2-inst eq for phi on cyl2}. Contrary to the two cases above, a
solution of \eqref{coccyl:insteq2} on the submanifold $\{y=y_0= $ const$\}$ does not extend trivially to a solution on the
full eight-dimensional space. This can be accomplished by choosing instead the parametrization $z = s(1-\ii t)$.
For $s$-independent $\phi$ the instanton equation becomes
 \begin{equation}\label{cocyl:insteqalt}
   \frac 12 \big(\ii - t \big) \frac {\dd \phi}{\dd t} = -\phi + \overline \phi^2,
 \end{equation} 
 and the slices $\{s = s_0=$ const$\}$ carry the cylinder metric $s_0^2 ( \dd t^2 + g^{(6)})$. Both \eqref{coccyl:insteq2} and \eqref{cocyl:insteqalt} can be obtained from the Spin(7)-instanton equation restricted to the cylinder, as in \eqref{G2-inst eq from Spin7}. In the first case one obtains exactly the cylindrical $G_2$-instanton equation by imposing $\nu_8 \lrcorner \F=0$, where $\nu_8$ is the 1-form normal to the cylinder. This condition is not satisfied for solutions of \eqref{cocyl:insteqalt} however, so that these do not solve the $G_2$-instanton equation. Substituting $t = \cot (\theta)$ in \eqref{cocyl:insteqalt} brings us back to the sine-cone equation \eqref{G2-inst eq for phi on sine-cone}.
\end{itemize}
 Of course, other foliations of $\Mm8$ are possible, and they lead to additional instanton equations. 
 
\begin{figure}[h!]
\begin{center}
\begin{tikzpicture}[>=stealth, domain=0:3.14]    
 \draw[->] (0,-1.5) -- (0,1.5) node[anchor=east] {$x$};    
 \draw[->] (-0.5,0) -- (3,0) node[anchor=north] {$y$};      
 \draw[color=blue, thick] (1,-1.5) -- (1,1.5); 
 \draw[color=red, thick] plot({sin(\x r)},{-cos(\x r)}); 
 \draw[thick] (0,1) -- (3,1); 
 \put(2,0.5){y};
\end{tikzpicture} \caption{The complex $z$-plane, $z=y-\ii x$. \Mm8 is a twisted product of \Mm6 with the right halfplane $\{y>0\}$. Embedded into \Mm8 are the sine-cone (red half-circle), cylinder (vertical blue line), and cone (horizontal black line). \Mm8 is foliated either by cylinders or cones, corresponding to the foliation of the half-plane by translations of the black and blue lines. A foliation by sine-cones is obtained through variation of the radius of the red half-circle. Upon a good parametrization of the three submanifolds the $G_2$-instanton equation on one of them becomes invariant under these shifts, so that a solution on a submanifold trivially extends to a Spin(7)-instanton on all of \Mm8.}\label{fig:submanifolds}
\end{center}
\end{figure}
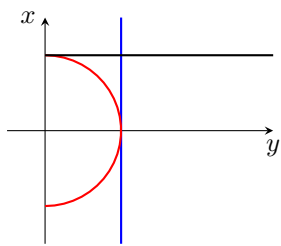

\paragraph{Yang-Mills actions.}
 The second-order equations obeyed by the solutions to the instanton equations admit a classical-mechanics interpretation in terms of a
particle moving in the plane. It can be obtained either by inserting the ansatz \eqref{ansatz1 for A} into the
Yang-Mills-Chern-Simons action \eqref{YM-action w torsion} or by differentiating the first-order equations. 
 Up to a common prefactor of $-12\,\Vol(\Mm6)$, the action for $\phi$ becomes
\begin{align}
\label{YM phi action cyl1}
 S^{cyl}_1 &= \int_\RR \left\{ \tfrac12 \phi^\prime\overline{\phi^\prime} 
                                      + V_3(\phi,\overline\phi) \right\} \dd \tau \:
\intertext{on the conformally parallel cylinder and}
 S^{cyl}_2 &=  \int_\RR \left\{ \tfrac12 \phi^\prime\overline{\phi^\prime} 
                                      + \Im(\phi\overline{\phi^\prime})
                                      + V_1(\phi,\overline\phi) \right\} \dd\tau 
\intertext{on the cocalibrated cylinder. On the metric cone and the sine-cone over \Mm6 the action reads}
 S^c &= \int_0^\infty \left\{ \tfrac12 y^4 \phi^\prime\overline{\phi^\prime}
                                      +y^2 \,V_0(\phi,\overline\phi) \right\} \dd y
\intertext{and}
\label{YM phi action sine-cone}
 S^{sc} &=  \int_0^\pi \left\{\tfrac12 \sin(\theta)^4 \phi^\prime\overline{\phi^\prime}
                           -2\,\sin(\theta)^3\,
                               \Re\big[(-\phi+\overline\phi^2)\ee^{\ii\theta}\overline{\phi^\prime}\big]
                           +\sin(\theta)^2\,V_3(\phi,\overline\phi) \right\} \dd\theta \:, 
\end{align}
respectively. Here, the potential $V_\kappa(\phi)$ is given by
\begin{align}
 V_\kappa(\phi,\overline\phi) &=  (\kappa{-}1)\,\phi\overline\phi 
                                  -(\tfrac\kappa3{+}1)\,(\phi^3+\overline\phi^3) 
                                  +2\,(\phi\overline\phi)^2 
\qquad\text{for}\quad \kappa=0,1\ \text{or}\ 3				  
\end{align}
as appropriate, and it is plotted in Figure~\ref{Fig:potentialplot} below.
We obtain the reduced Yang-Mills equations
\begin{alignat}{2}
\label{eom cyl1}
 \frac{\dd^2\phi}{\dd \tau^2} 
     &= 4\,\phi-12\,\overline\phi^2+8\,\phi^2\overline\phi \:,
    &&\qquad\text{{(conf. parallel cylinder)}} \\
 \frac{\dd^2\phi}{\dd \tau^2} 
     &= 2\ii\overline{\frac{\dd\phi}{\dd \tau} }-8\,\overline\phi^2+8\,\phi^2\overline\phi \:,
    &&\qquad\text{{(cocalibrated cylinder)}} \\
 y^2\,\frac{\dd^2\phi}{\dd y^2} 
    &=  -4\,y\frac{\dd\phi}{\dd y}  -2\phi-6\,\overline\phi^2+8\,\phi^2\overline\phi \:,
    &&\qquad\text{{(metric cone)}} \\
\label{eom sine-cone}
 \sin(\theta)^2\,\frac{\dd^2\phi}{\dd \theta^2} 
    &= -4\sin(\theta)\,\ee^{\ii\theta}\,\Big(\frac{\dd \phi}{\dd \theta} +8\ii(-\phi+\overline\phi^2)\Big)
       -2\phi-6\,\overline\phi^2+8\,\phi^2\overline\phi \:.
    &&\qquad\text{{(sine-cone)}} 
\end{alignat} 
The actions \eqref{YM phi action cyl1}-\eqref{YM phi action sine-cone} and equations of motion \eqref{eom cyl1}-\eqref{eom sine-cone} can be interpreted as describing the motion of a particle in the plane under the influence of a potential $-V_\kappa$ with one of the following effects:
\begin{itemize}
\item The equation of motion on the cocalibrated cylinder contains a term proportional to $\ii\overline{\phi^\prime}$, which mimics the Lorentz force exerted on the particle by a magnetic field perpendicular to the $\phi$-plane.
\item On the metric cone a friction term $-\phi^\prime$ appears in the equation.
\item On the sine-cone both effects appear. The friction coefficient even becomes negative for $\phi>\pi/2$, giving rise to a velocity dependent accelerating force on the particle. Moreover, the particle mass is time-dependent. 
\end{itemize}

\begin{figure}
\centerline{
        \begin{overpic}[scale=.27]%
            {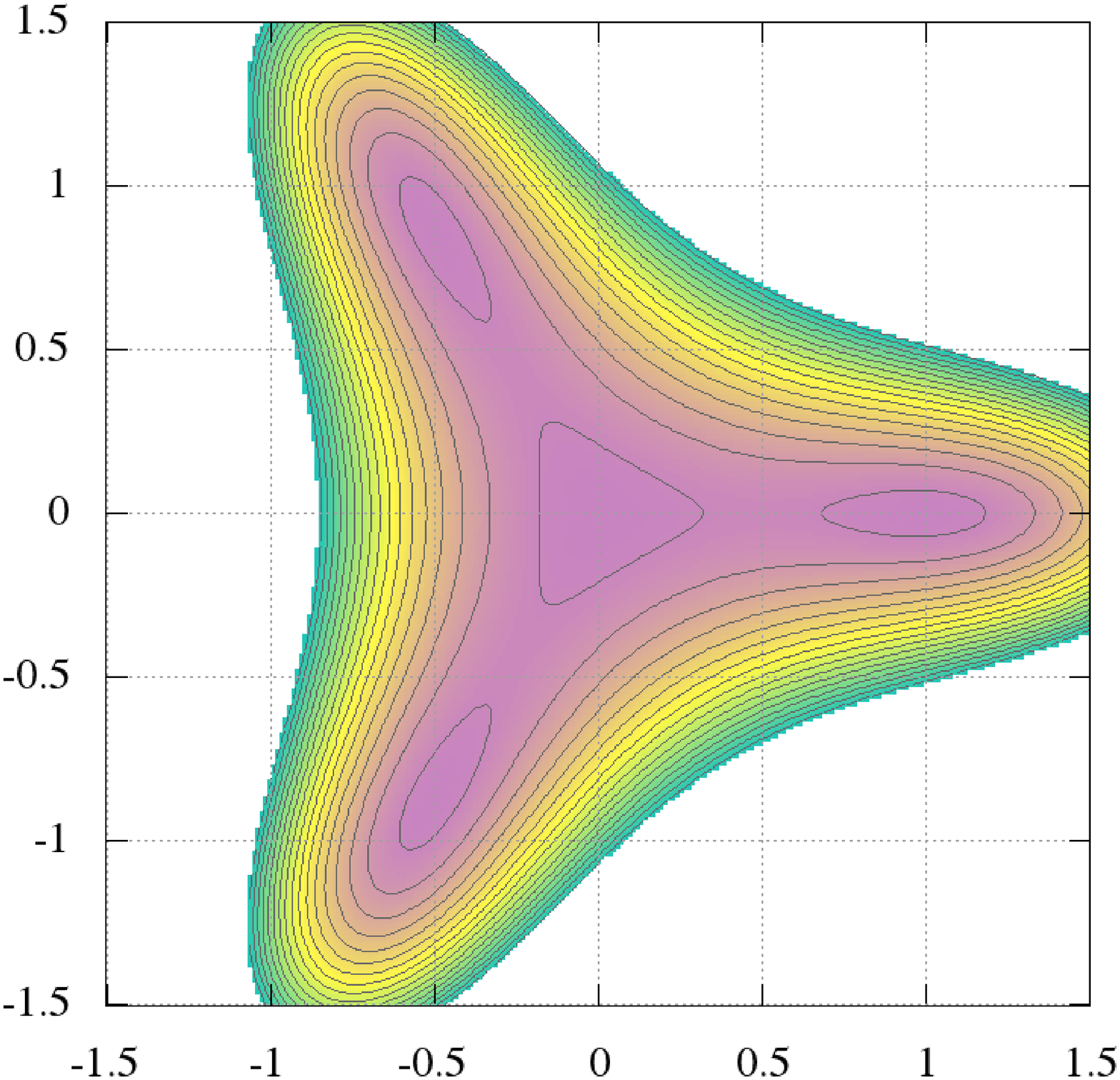} \put(150,165){\large $\kappa=3$}
      \end{overpic}
\hfill
      \begin{overpic}[scale=.27]%
            {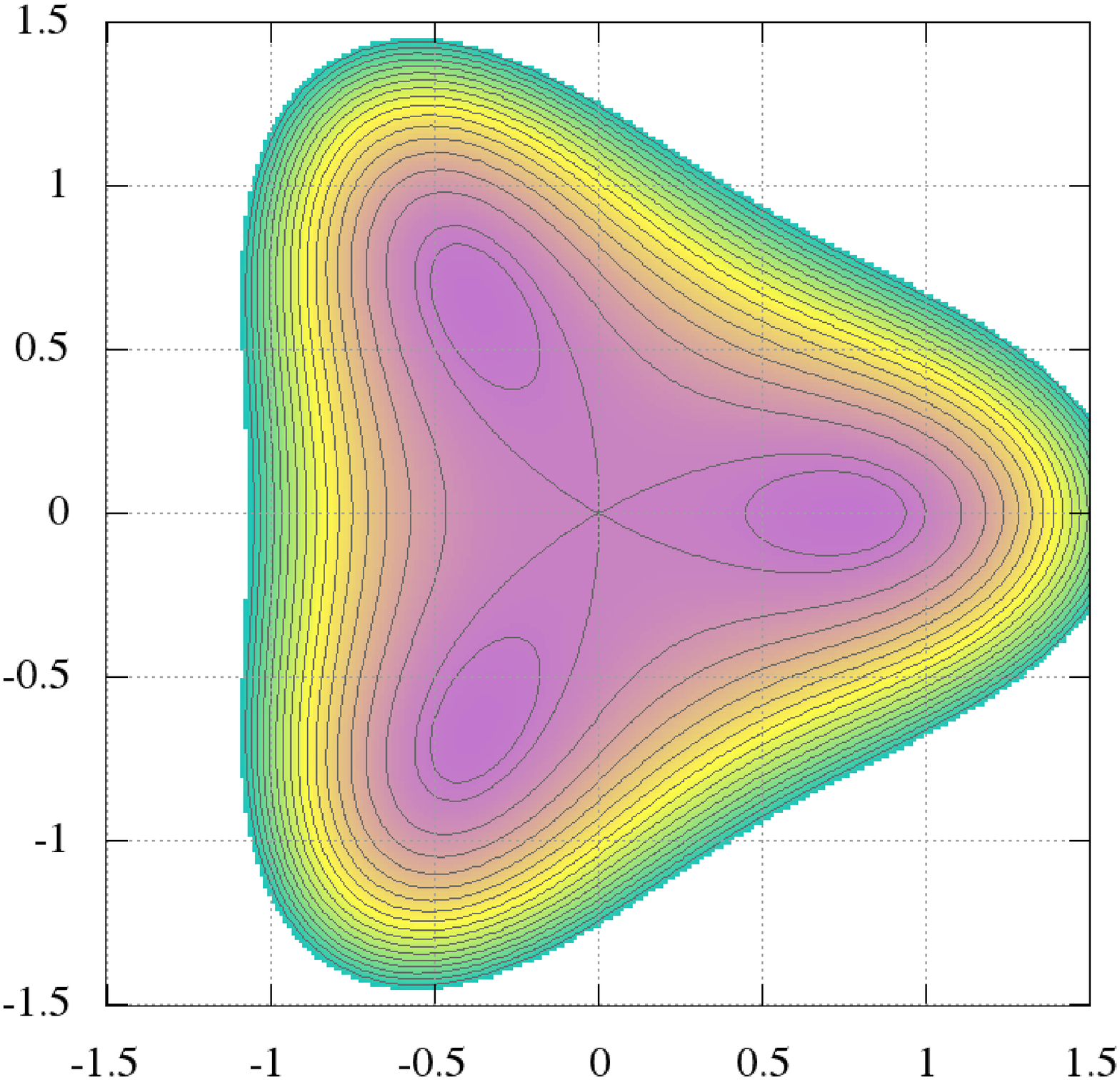} \put(150,165){\large $\kappa=1$}
      \end{overpic}
}$\:$\\
\centerline{
      \begin{overpic}[scale=.27]%
            {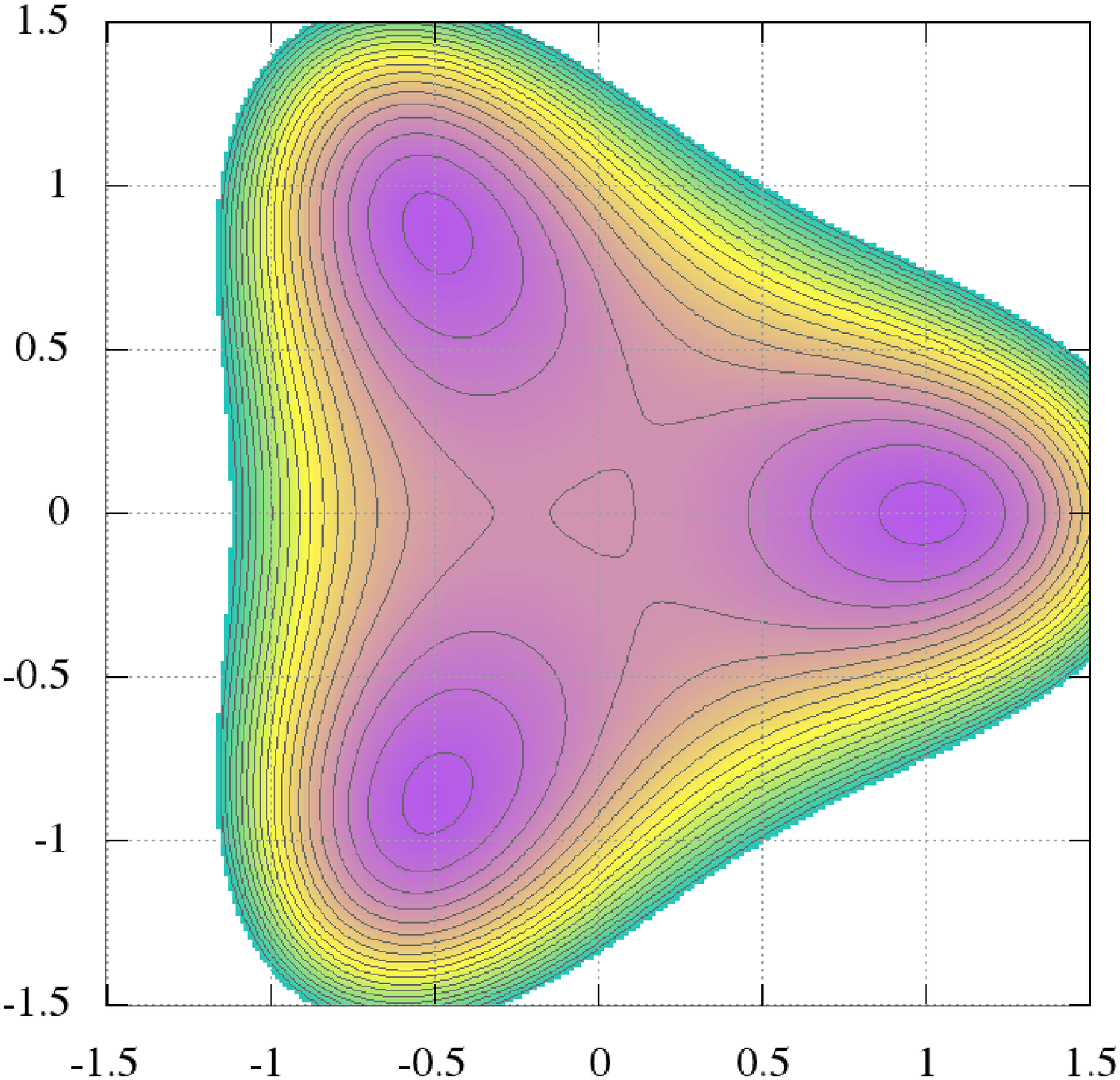} \put(150,165){\large $\kappa=0$}
      \end{overpic}
}
\caption{Contour plots of the potential $-V_\kappa(\phi)$ for $\kappa=3$ (top left), $\kappa=1$ (top right) and $\kappa=0$ (bottom). In all three cases, $-V_\kappa(\phi)$ has local maxima at $\phi=1$ and $\phi=\exp(\pm2\pi\ii /3)$. Moreover, at $\phi=0$, the potential $-V_3(\phi)$ has an additional maximum on the same level as the other three maxima, $-V_1(\phi)$ has a saddle point, and $-V_0(\phi)$ has a local minimum. The friction term in the equation of motion on the cone (which has $\kappa=0$) leads to solutions going from one of the maxima to the minimum at $\phi=0$. For $\kappa=1$, i.e. the cocalibrated cylinder, our instanton solutions interpolate between the three maxima $\phi=1,\exp(\pm 2\pi \ii/3)$, and the same is true for the sine-cone, with $\kappa=3$. However, $\kappa=3$ also covers the conformally parallel cylinder, which only admits solutions between $\phi=0$ and one of the three other maxima.}\label{Fig:potentialplot}
\end{figure}

\subsection{Solutions for gauge group $G$}\label{subsec_solutions}
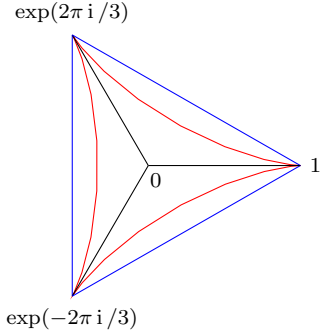
\begin{figure}[h!]
\begin{center}
\begin{tikzpicture}[domain=0:9.42]    
  \draw[color=red] plot ({2*cos(\x r)*cos(\x/3 r) + 2/3*sin(\x r)*sin(\x/3 r)},{2*cos(\x r)*sin(\x/3 r) -2/3*sin(\x r)*cos(\x/3 r)  });
   \draw[color=blue](-1,-1.732) -- (-1,1.732);
   \draw(-1,-1.732) -- (0,0);
   \draw[color=blue](-1,-1.732) -- (2,0);
   \draw(-1,1.732) -- (0,0);
   \draw[color=blue](-1,1.732) -- (2,0);
  \draw(0,0) -- (2,0);
 \node[below=0.8] at (-1,-1.732) {\footnotesize $\exp(-2\pi\, \ii/3)$};
\node[above=0.8] at (-1,1.732) {\footnotesize $\exp(2\pi\, \ii/3)$};
 \node at (2.2,0) {\footnotesize $1$};
  \node at (0.1,-0.2) {\footnotesize $0$};
\end{tikzpicture} \caption{Instantons in the complex $\phi$ plane. The nodes
correspond to the four SU(3)-instantons on the nearly K\"ahler manifold $\Mm6$, whereas the edges are interpolating Spin(7)-instantons on $\Mm8=Cyl(C(\Mm6))$ or submanifolds thereof. The blue edges can be realized as $G_2$-instantons on the cocalibrated cylinder $Cyl(\Mm6)$, the red ones solve the nearly parallel $G_2$-instanton equation on the sine-cone $C_s(\Mm6)$,
 and the black edges are solutions on the cone $C(\Mm6)$. The 3-symmetry of the nearly K\"ahler manifold is reflected in the permutation symmetry of the diagram.}\label{diagram_instantons}
\end{center}
\end{figure}

 Here we collect the known finite-action solutions to the instanton equations \eqref{real de for phi}.
 The conformally parallel cylinder, and thus the cone, was discussed in \cite{Bauer2010,Harland2009,Ivanova2009,Harland2010}:
\begin{align}
 \label{phi kink}
 \phi(\tau) = \frac{c}{2}\Big( 1 - \tanh(\tau{-}\tau_0)\Big)
 \qquad\text{with}\quad
 c=1 \quad\text{or}\quad c=\exp\bigl(\pm\ii\sfrac{2\pi}{3}\bigr)\:.
\end{align}
The solution
\eqref{phi kink} interpolates between the stationary SU(3)-instantons $\phi\rightarrow c$ for
$\tau\rightarrow-\infty$ and $\phi\rightarrow 0$ for $\tau\rightarrow\infty$, and it is represented by the black edges in
Figure~\ref{diagram_instantons}. Solutions to \eqref{G2-inst eq for phi on cyl2} on
the cocalibrated cylinder have been found in \cite{Harland2009} and are given by
 \begin{equation}
   \label{phi modified kink}
 \phi(\tau) = -\frac{c}{2}\Big( 1+ {\ii \sqrt 3} \tanh\big[{\sqrt 3} (\tau{-}\tau_0)\big]\Big)\:,
 \end{equation} 
 with $c$ as above being equal to one of the three non-trivial fixed points. These solutions interpolate between the fixed points $c\cdot \exp(-2\pi \ii/3)$ and $c\cdot \exp(2\pi \ii/3)$, as is illustrated by the blue edges in Figure~\ref{diagram_instantons}. We also found solutions for the sine-cone equation \eqref{G2-inst eq for phi on sine-cone}, namely
 \begin{equation}\label{sineconekink}
   \phi(\theta) = c\,\Big(\cos(\theta) -\sfrac \ii 3\sin(\theta )\Big)\,\ee^{\ii \theta/3}\:.
 \end{equation} 
 These are drawn in red in Figure~\ref{diagram_instantons} and interpolate between $c$ for $\theta\rightarrow 0$ and $c\cdot \exp(-2\pi \ii/3)$ for $\theta \rightarrow \pi$. Moreover, they give rise to solutions of the Spin(7)-instanton equation \eqref{cocyl:insteqalt} restricted to the cylinder, upon substituting $\theta = $ arccot$(t)$. Explicitly, we get
 \begin{equation}
   \phi(t) = c\,\frac{(t-\frac \ii 3)(t+\ii)^{1/3}}{ (t^2 +1)^{2/3}}\:.
 \end{equation} 

\subsection{Gauge groups $G_2$ and Spin(7)}

  The ansatz for the gauge field chosen above relies on the fact that  
on a reductive homogeneous space $\M=G/H$ with Lie-algebra
splitting $\mathfrak g = \mathfrak h \oplus \mathfrak m$ we can  
identify the tangent space $T_x\M$ with the
$H$-module~$\mathfrak m$. Stated more globally, we have an isomorphism  
of vector bundles $T\M = G \times_H
\mathfrak m$, and $H$ plays the role of the structure group of $T\M$.  
On the other hand, the nearly K\"ahler spaces carry an SU(3)-structure  
with corresponding SU(3)-frame bundle $P$, and we can also identify  
$T\M = P\times_{SU(3)} \mathfrak m$. In fact, $H$ is a subgroup of  
SU(3) for nearly K\"ahler manifolds, and $\mathfrak m$ carries a  
SU(3)-module structure which upon restriction to $H$ gives back the  
action of $H$ on $\mathfrak m$. Hence, we meet the two reductive decompositions
  \begin{equation}
    \mathfrak g = \mathfrak h\oplus \mathfrak m\und \mathfrak g_2 =  
\mathfrak {su}(3) \oplus \mathfrak m\:,
  \end{equation}
  and thus we can view $\mathfrak m$ either as a  
subspace of $\mathfrak g$ or of $\mathfrak g_2$. In the previous subsection  
we constructed several gauge fields on $Cyl(C(\Mm6))$, the cylinder  
over the cone over a nearly K\"ahler six-manifold, which contain terms of  
the form $e^a\Phi_{ab} I_b$. Locally, the $I_a$ are elements of  
$\mathfrak m$, considered as a subspace of $\mathfrak g$ so far.  
According to the above discussion we can also view them as elements of  
$\mathfrak g_2$ and thus obtain connections with gauge group $G_2$. 
Since the instanton equations remain unchanged, this is merely a  
reinterpretation of the earlier results. In the case of $S^6 =  
G_2/$SU(3) nothing changes at all. \\

  This new interpretation has some nice features however. First of all, it is
  valid not only for homogeneous spaces but also, for instance, on  
nearly K\"ahler manifolds obtained as sine-cones over Sasaki-Einstein  
5-manifolds \cite{FIMU06}.
Additionally, it provides a geometric interpretation for some of the  
instantons. Both nearly K\"ahler and nearly parallel $G_2$-manifolds  
possess a so-called canonical connection, which has 
holonomy SU(3) or $G_2$, respectively, has totally skew-symmetric torsion
and satisfies the instanton equation. 
On manifolds with reduced holonomy group $G_2$ or  
Spin(7) on the other
hand, the Levi-Civita connection provides one particular solution to  
the corresponding instanton equation. \\

  Consider again Figure~\ref{diagr of geometries}. The four  
spaces with reduced structure group SU(3), $G_2$ or Spin(7) are related by 
certain geometric operations, and each carries a distinguished instanton.
 As a gauge field, the
Levi-Civita connection on a Riemannian manifold can be identified with  
the one on its cylinder, because their connection
1-forms (or gauge fields) are the same. Thus we end up with four  
geometries giving rise to three different instantons.
They correspond to the canonical connection on the base $\M$ for $\phi=0$,
the Levi-Civita connection on the cone $C(\M)$ for $\phi=1$, and
additionally the canonical $G_2$-connection on the sine-cone  
$C_s(\M)$, which is gauge-equivalent to the non-stationary solution  
\eqref{sineconekink}. The solution \eqref{phi kink} interpolating  
between $\phi=0$ and $\phi=1$ therefore related the pull-back of the  
canonical SU(3)-connection on \Mm6 with the Levi-Civita connection on  
the cone $C(\Mm6)$. \\

In the special case $\Mm6 = S^6$ the cone is simply $\mathbb R^7\setminus\{0\}$, and  
\eqref{phi kink} reproduces the octonionic instanton  
\cite{Guenaydin}. This follows from the fact that our ansatz gives the  
most general $G_2$-invariant gauge field on $\mathbb R^7$, while
the octonionic instanton is $G_2$-invariant as well.\footnote{
We thank Derek Harland for pointing out this argument.} 
In the Appendix 
we verify explicitly the coincidence of the curvature tensors.\\

  To make the identification of the gauge field for $\phi=1$ with the  
Levi-Civita connection on the cone more precise, we consider the  
following explicit realization of the $\mathfrak m$-generators $I_a$  
as elements of $\mathfrak g_2 \subset \mathfrak {so}(7)$. Let indices  
$a,b,c$ run from 1 to 6, and define skew-symmetric $7\times  
7$-matrices $I_a$ by
  \begin{equation}\label{IaforConeLC}
    (I_a) _{b7}  =\delta_{ab} \und (I_a)_{bc}  = \sfrac 12  
(\text{Re}\,\Omega)_{abc}\:.
  \end{equation}
  The cone metric can be written as $\ee^{2\tau}(d\tau^2 + g^6)$, where  
$\tau$ is the logarithm of the radial coordinate, $y=\ee^\tau$. In the  
orthonormal frame $\{\ee^{\tau}e^a$, $\ee^\tau d\tau\}$ its Levi-Civita  
connection assumes the form
  \begin{equation}\label{G2ConeLC}
    \Gamma = \hat A^6 + e^a I_a\:,
  \end{equation}
  where $\hat A^6= e^iI_i$ is the canonical connection of the base and  
the $I_a$ are given by the matrices \eqref{IaforConeLC} acting on the  
tangent space. \\

  The canonical connection $\hat A$ on a nearly K\"ahler or nearly  
parallel $G_2$-manifold is obtained from its Levi-Civita connection by  
adding a suitable multiple of the canonical 3-form Re$(\Omega)$ or  
$\Psi$, respectively \cite{Friedrich02}. For the sine-cone over a nearly K\"ahler  
manifold this recipe leads to the following expression:
  \begin{equation}
    \hat A^{7} = \hat A^6 + \big[ \cos(\theta)\delta_{ab} - \sfrac 13  
\sin(\theta)\omega_{ab}\big] e^a I_b - \sfrac 13 \dd\theta \,J\:,
  \end{equation}
  where $J$ is the almost complex structure acting non-trivially only  
on the tangent space to $\Mm6$ and the $I_a$ are skew-symmetric  
$7\times 7$-matrices
  \begin{equation}
           (I_a)_{b7} = \delta_{ab} \und (I_a)_{bc} = \sfrac 12  
\text{Re}(\ee^{\ii\theta}\Omega)_{abc}\:.
  \end{equation}
   Again, indices $a,b,c$ run from 1 to 6. Upon a gauge transformation
\begin{equation}
   \big(\dd+\hat A^7\big)\ \mapsto\ \ee^{- \theta J/3} \big(\dd+\hat A^7  
\big)\,\ee^{\theta  J/3}\:,
\end{equation}
  the connection form is transferred to the form of our ansatz  
\eqref{ansatz1 for A}, with $\phi$ given by \eqref{sineconekink} and  
the $I_a$ by \eqref{IaforConeLC}.\\

  Now we turn to gauge group Spin(7). Similarly to the decomposition  
$\mathfrak g_2 = \mathfrak{su}(3) \oplus \mathfrak m$ we have a  
decomposition
  \begin{equation}
    \mathfrak{spin}(7) =   \mathfrak g_2 \oplus \mathfrak m'\:,
  \end{equation}
  where $\mathfrak m'$ is the seven-dimensional irreducible $\mathfrak  
g_2$-module. We can introduce a gauge field in the form
  \begin{equation}
    A = \hat A^7 + \phi\, \hat e^a  \hat I_a\:,
  \end{equation}
  where $\hat I_a$ are the generators of $\mathfrak m'$ and $\hat e^ a$  
the dual 1-forms. As before, $\hat A^ 7$ denotes the canonical  
$G_2$-connection on the sine-cone $C_s(\Mm6)$; it can be written as  
$\hat A^ 7 =  \hat e^i \hat I_i$ for generators $\hat I_i$ of  
$\mathfrak g_2$. The Spin(7)-instanton equation for $\phi$ turns out  
to coincide with \eqref{G2-inst eq for phi on cyl1} if we substitute  
$r=\ee^\tau$, and therefore a solution is given by
  \begin{equation}\label{spin(7)kink}
    \phi(\tau ) = \sfrac 12\big(1-\tanh(\tau{-}\tau_0)\big)\:.
  \end{equation}
  The gauge field interpolates between the canonical nearly parallel  
$G_2$-connection and the Levi-Civita connection on the cone. For $\Mm6  
= S^6$ we have $\Mm8=\mathbb R^8\setminus \mathbb R$, and \eqref{spin(7)kink} gives rise  
to the octonionic instanton \cite{Fairlie84, Fubini85}.

\section{Conclusions and outlook}

For a given nearly K\"ahler manifold $\Mm6$, there are several interesting related
geometries in dimension seven and eight with $G_2$- and Spin(7)-structures, respectively.
They give rise to different instanton equations, and we have presented
finite-action solutions to all of them, mostly by collecting
earlier results. A unified
approach to these equations has been developed by embedding all these seven-manifolds
into the Spin(7)-manifold $\Mm8= C(C_s(\Mm6)
= Cyl(C(\Mm6))$. The eight-dimensional Spin(7)-instanton equation \eqref{cplx de for phi}
induces the
respective $G_2$-instanton equations upon restriction to the cone or the sine-cone over
\Mm6, but assumes a somewhat more general form on the cylinder, which we have been able
to solve as well.\\

Ultimately we would like to understand the moduli spaces of $G$-invariant instantons on 
$\Mm8$ built over an arbitrary homogeneous space $\Mm6=G/H$, and for any fixed gauge group. The results
presented here can be understood as a first step in this direction, by restricting
attention to
instantons that are invariant under translations in one direction. Several non-trivial
solutions of this type exist, and
the full moduli space is certainly much larger. \\

Among the instantons presented here are the octonionic instantons on $\mathbb R^7$ and
$\mathbb R^8$. It is well-known
that they can be lifted to solutions of heterotic supergravity, and in fact this is true
for the $G_2$- and Spin(7)-instantons \eqref{phi kink} and \eqref{spin(7)kink} for an arbitrary nearly K\"ahler base
manifold \cite{Harland11}. An
interesting question arises whether this is also possible for the instantons on the
coclibrated cylinder and the
sine-cone. \\

Most investigations of moduli spaces of heterotic string vacua so far have focussed on
solutions without
fluxes, based on integrable geometries like Calabi-Yaus, $G_2$- or Spin(7)-manifolds. The
gauge field is normally
required to coincide with the Levi-Civita connection in these cases. Our investigation
shows that the gauge sector of
heterotic string theory on certain conical Spin(7)-manifolds admits several deformations
away from the Levi-Civita
connection. If these embed into string theory, the moduli space of integrable
backgrounds captures only a small
fraction of the full vacuum structure of heterotic string theory.\\

A construction similar to the one presented here for six-dimensional nearly K\"ahler
manifolds should be possible for
a five-dimensional Sasaki-Einstein manifold. In this case the cone is a Calabi-Yau
3-fold, the sine-cone is a
nearly K\"ahler manifold, and the cone over the sine-cone has $G_2$-holonomy
group\cite{FIMU06, Boyer2007}. Then one could go ahead
and consider also the sine-cone over the sine-cone, which is a nearly parallel
$G_2$-manifold, and so on. It turns out, however, that the
instanton equation on the cone over a Sasaki-Einstein manifold assumes a more complicated
form than is the case for
nearly K\"ahler manifolds, which could be an obstacle to obtaining explicit solutions
\cite{Harland11}.

\section*{Acknowledgements}

We thank Derek Harland for helpful comments. This work was supported in part by 
the cluster of excellence EXC 201 ``Quantum Engineering and Space-Time Research'',
by the Deutsche Forschungsgemeinschaft (DFG), by the 
Russian Foundation for Basic Research (grants RFBR 09-02-91347 and 10-01-00178) 
and by the Heisenberg-Landau program.

\newpage
\begin{appendix}

\section{Comparing the instanton on $\RR^7$ and 
on the cone over $G_2/\SU(3)$} \label{app:OctInstConeInst}
Since $G_2/\SU(3)$ with its nearly K\"ahler metric is the round six-sphere, the cone $C(G_2/\SU(3))$ gives $\RR^7$.  In this appendix we compare explicitly the octonionic instanton on $\RR^7$, constructed in \cite{Guenaydin}, with the instanton \eqref{phi kink} on the cone and show that they coincide.

\paragraph{$G_2$-structure on $\RR^7$ and SU(3)-structure on $S^6$.}
On $\RR^7$ we have a canonical integrable $G_2$-structure
\begin{align}
 \Psi = \sfrac16 f^\mathbb{O}_{\hat a\hat b\hat c} \dd x^{\hat a} \wedge \dd x^{\hat b} \wedge \dd x^{\hat c} \:,
\end{align}
with $x^{\hat a}$ ($\hat a=1,\ldots,7$) being Euclidean coordinates on $\RR^7$.  This $G_2$-structure induces an SU(3)-structure on $S^6$ defined by the forms
\begin{align}
 \omega &= \imath^*(\dd r \ins \Psi) \und \Omega = \imath^*(\dd r \ins *\Psi) + \ii \, \imath^* \Psi \:,
\end{align}
where $r^2=x^{\hat a}x^{\hat a}$ and $\imath:S^6\rightarrow\RR^7$ is the inclusion map of the six-sphere. The two- and three-forms $\omega$ and $\Omega$ define a nearly K\"ahler structure on $S^6$ \cite{Cabrera2006}. Since up to isometry there is only one nearly K\"ahler structure on $S^6$ \cite{Friedrich2005}, this has to be the same structure we considered on $G_2/SU(3)$.

\paragraph{Seven-dimensional representation of $G_2$.}
In \cite{Guenaydin} generators of $G_2$ in the seven-dimensional representation were constructed via the spinor representation of SO(7). The generators of $\SO(7)$ in the spinor representation are given in terms of the gamma matrices in seven dimensions by
\begin{align}
 \Gamma_{\hat a \hat b} := \Gamma_{[\hat a} \Gamma_{\hat b]} 
 \qquad\text{with}\quad \hat a,\hat b = 1,\ldots,7\:.
\end{align}
The subgroup $G_2 \subset \SO(7)$ is generated by the subset of generators $\{G_{\hat a\hat b}\}
\subset \{\Gamma_{\hat a\hat b}\}$ satisfying the constraints
\begin{align}
 f^{\mathbb{O}}_{\hat a\hat b\hat c} G_{\hat b\hat c} = 0 \:.
\end{align}
Their commutation relation inherited from SO(7) is
\begin{align}
\label{commutator G}
 [G_{\hat a\hat b},G_{\hat c\hat d}] =  2\,\delta_{\hat c[\hat b}\,G_{\hat a]\hat d}
                                       -2\,\delta_{\hat d[\hat b}\,G_{\hat a]\hat c}
                                       +\tfrac12\,\big( (*\Psi)_{\hat c\hat d\hat e[\hat a}G_{\hat b]\hat e}
                                                   -(*\Psi)_{\hat a\hat b\hat e[\hat c}G_{\hat d]\hat e}\big) \:,
\end{align}
and they are normalized such that
\begin{align}
\label{normalization G}
 \tr(G_{\hat a\hat b}G^{\hat c\hat d}) = -6\,\big[\delta\indices{_{\hat a}^{\hat c}}\delta\indices{_{\hat b}^{\hat d}}
                                             +\tfrac14 (*\Psi)\indices{_{\hat a\hat b}^{\hat c\hat d}}\big] \:.
\end{align}

\paragraph{Instanton solution on $\RR^7$.}
The instanton solution constructed in \cite{Guenaydin} reads 
\begin{align}
\label{NG connection}
 \A &= G_{\hat a\hat b} f_{\hat b}(x)\: \dd x^{\hat a}\qquad\longrightarrow\\[2pt]
\label{NG curvature}
 \F &= \big[ 2\,f_{\hat c[\hat a} G_{\hat b]\hat c} 
             -2\,f_{\hat c} G_{\hat c[\hat a} f_{\hat b]}
             -G_{\hat a\hat b} f_{\hat c} f_{\hat c}
             -\tfrac12 (*\Psi)_{\hat a\hat b\hat c\hat d}f_{\hat e}G_{\hat e\hat c}f_{\hat d} \big]\,
           \dd x^{\hat a} \wedge \dd x^{\hat b} \:,
\intertext{with}
  f_{\hat a}  &= \partial_{\hat a} f  \quad\text{and}\quad
  f_{\hat a\hat b}  = \partial_{\hat a} \partial_{\hat b} f \qquad\text{for}\qquad
  f(y)   = - \tfrac{1}{2} \log (\rho^2 + r^2) \:,
\end{align}
where $\rho$ is an arbitrary scale parameter.  In order to compare this solution with the solutions on the cone over $G_2/\SU(3)$, it is necessary to chose an orthonormal coframe $\{e^{\hat a}\}$ such that $e^7=\dd r$ and the one-forms $e^a$ are the left-invariant one-forms defined in Section \ref{NK section}.  We consider SU(3) to be embedded in $G_2$ such that it is the stabilizer of $\dd r$.  This implies that we can write the matrices $G_{\hat a\hat b}$ in this frame as
\begin{align}
\label{appendixA G_ab <-> I_A}
 G_{ab} = \sfrac 12 \sqrt{3} f_{abc} \check I_c + \sqrt{3} f_{abi} \check I_i 
 \und
 G_{a7} = \sqrt{3} \delta_{ab} \check I_b \:,
\end{align}
where $\check I_i$ and $\check I_a$ are the generators of $\su(3)$ and $\su(3)^\perp$, respectively, in the seven-dimensional representation.  The matrices defined by \eqref{appendixA G_ab <-> I_A} satisfy the commutation relation \eqref{commutator G} and are compatible with \eqref{normalization G} if the generators $\check I_A$ are normalized such that
\begin{align}
 \tr(\check I_A\check I_B) = - \delta_{AB} \:.
\end{align}
The curvature \eqref{NG curvature} written in this frame reads
\begin{equation} \label{F_R7}
F_{a7} = -\frac{2 \sqrt{3} \, \rho^2}{(\rho^2+r^2)^2} \check I_a \und
F_{ab} = \frac{\sqrt{3}}{(\rho^2 +r^2)^2} 
  \left[ -(2\rho^2+r^2) f_{abi}\check I_i-\rho^2f_{abc} \check I_c \right]\:,
\end{equation}
and it coincides with the one of the connection \eqref{phi kink} for $c=1$.
This completes the proof.

\end{appendix}

\end{document}